\overfullrule=0pt
\input harvmac

\def\sffb{{{\sqrt{f}\over\sqrt{\fb}}}}
\def\sfbf{{{\sqrt{\fb}\over\sqrt{f}}}}
\def\p{{\partial}}
\def\bp{{\overline\partial}}
\def\a{{\alpha}}
\def\b{{\beta}}
\def\g{{\gamma}}

\def\half {{1 \over 2}}
\def\k{{\kappa}}
\def\kh{{\widehat k}}
\def\bh{{\widehat b}}
\def\kb{{\overline\kappa}}
\def\nablab{{\overline\nabla}}
\def\S{{\Sigma}}
\def\d{{\delta}}
\def\T{{\Theta}}
\def\Tb{{ \overline\Theta}}
\def\Xb{{ \overline X}}
\def\Nb{{ \overline N}}
\def\Rb{{ \overline R}}
\def\Phib{{ \overline \Phi}}
\def\Mb{{ \overline M}}
\def\fb{{ \overline f}}
\def\ab{{ \overline a}}
\def\mb{{ \overline m}}
\def\Db{{ \overline D}}
\def\Wb{{ \overline W}}
\def\Gb{{ \overline G}}
\def\xb{{ \overline x}}
\def\sb{{ \overline s}}
\def\pb{{ \overline p}}
\def\lb{{ \overline l}}
\def\nb{{ \overline n}}
\def\ttb{{ \overline t}}
\def\t{{\theta}}
\def\tb{{\overline \theta}}

\def\l{{\lambda}}
\def\bl{{\overline\lambda}}
\def\wb{{\overline w}}

\lref\witten{N. Berkovits, C. Vafa and E. Witten, {\it
Conformal Field Theory of AdS Background with Ramond-Ramond Flux},
JHEP 9903 (1999) 018, hep-th/9902098.}
\lref\gaume{L. Alvarez-Gaum\'e and D.Z. Freedman, {\it Kahler Geometry
and the Renormalization of Supersymmetric $\sigma$ Models}, Phys. Rev.
D22 (1980) 846.}
\lref\grizanon{M. Grisaru, A. van
de Ven and D. Zanon, {\it Two-Dimensional Supersymmetric Sigma Models
on Ricci Flat Kahler Manifolds are Not Finite}, Nucl. Phys. B277
(1986) 388.}.
\lref\fms{D. Friedan, E. Martinec and S. Shenker, 
{\it Conformal Invariance, Supersymmetry and String Theory},
Nucl. Phys. B271 (1986) 93.} 
\lref\maluf{N. Berkovits and J. Maldacena, {\it N=2 Superconformal
Description of Superstring in Ramond-Ramond Plane Wave Backgrounds},
hep-th/0208092.}
\lref\pures{N. Berkovits, {\it 
Super-Poincar\'e Covariant Quantization of the
Superstring}, JHEP 0004 (2000) 018, hep-th/0001035.} 
\lref\meads{N. Berkovits and O. Chand\'{\i}a, {\it Superstring Vertex
Operators in an $AdS_5\times S^5$ Background}, Nucl. Phys. B596 (2001) 185,
hep-th/0009168.}
\lref\mone{J.M. Maldacena,
{\it The Large N Limit of Superconformal Field Theories and Supergravity},
Adv. Theor. Math. Phys. 2 (1998) 231, Int. J. Theor. Phys. 38 (1999) 1113,
hep-th/9711200\semi S.S. Gubser, I.R. Klebanov and A.M. Polyakov,
{\it Gauge Theory Correlators from Non-Critical String Theory},
Phys. Lett. B428 (1998) 105, hep-th/9802109
\semi E. Witten, {\it Anti-deSitter Space
and Holography}, Adv. Theor. Math. Phys. 2 (1998) 253, hep-th/9802150.}
\lref\confops{N. Berkovits, {\it Conformal Field Theory for the
Superstring in a Ramond-Ramond Plane Wave Background}, JHEP 0204
(2002) 037, hep-th/0203248.}
\lref\guk{N. Berkovits, S. Gukov and B.C. Vallilo, {\it Superstrings
in 2D Backgrounds with RR Flux and New Extremal Black Holes},
Nucl. Phys. B614 (2001) 195, hep-th/0107140.}
\lref\gris{P.S. Howe and G. Papadopoulos, {\it N=2 D=2
Supergeometry}, Class. Quant. Grav. 4 (1987) 11\semi
S.J. Gates Jr., L. Liu and N. Oerter, {\it Simplified SU(2)
Spinning String Superspace Supergravity}, Phys. Lett. B214 (1988) 26\semi
M. Grisaru and M.E. Wehlau, {\it Prepotentials for
(2,2) Supergravity}, Int. J. Mod. Phys. A10 (1995) 753, hep-th/9409043.}
\lref\topo{N. Berkovits,
{\it The Ten-Dimensional Green-Schwarz Superstring is a Twisted 
Neveu-Schwarz-Ramond String}, Nucl. Phys. B420 (1994) 332\semi
N. Berkovits and C. Vafa, {\it On the Uniqueness of String Theory},
Mod. Phys. Lett. A9 (1994) 653, hep-th/9310170\semi
N. Berkovits and C. Vafa, {\it N=4 Topological Strings}, Nucl. Phys.
B433 (1995) 123, hep-th/9407190.}
\lref\ufourp{N. Berkovits, {\it
The Heterotic Green-Schwarz Superstring on an N=(2,0) Worldsheet},
Nucl. Phys. B379 (1992) 96, hep-th/9201004\semi
N. Berkovits, {\it Calculation of Green-Schwarz Superstring Amplitudes
using the N=2 Twistor-String Formalism}, Nucl. Phys. B395 (1993) 77,
hep-th/9208035.}
\lref\tseyt{A.A. Tseytlin, {\it Semiclassical Quantization of 
Superstrings: $AdS_5\times S^5$ and Beyond}, hep-th/0209116.}
\lref\russotseytlin{J.G. Russo and A.A. Tseytlin, {\it A Class of Exact
PP Wave String Models with Interacting Light Cone Gauge Actions},
JHEP 0209 (2002) 035, hep-th/0208114.}
\lref\iiauf{G. Bonelli, {\it Matrix String Models for Exact (2,2)
String Theories in R-R Backgrounds}, hep-th/0209225.}
\lref\metsaev{R.R. Metsaev, {\it Type II Green-Schwarz 
Superstring in Plane Wave Ramond-Ramond Background}, Nucl. Phys. B625
(2002) 70, hep-th/0112044.}
\lref\bmn{D. Berenstein, J. Maldacena and H. Nastase, {\it
Strings in Flat Space and PP Waves from N=4 Super-Yang-Mills},
JHEP 0204 (2002) 013,
hep-th/0202021.}
\lref\maoz{J. Maldacena and L. Maoz, {\it Strings on PP Waves and
Massive Two-Dimensional Field Theories}, hep-th/0207284.}

\Title{\vbox{\hbox{IFT-P.074/2002 }}}
{\vbox{
\centerline{\bf N=2 Sigma Models for Ramond-Ramond Backgrounds}}}
\bigskip
\centerline{Nathan Berkovits\foot{e-mail: nberkovi@ift.unesp.br}}
\smallskip
\centerline{\it Instituto de F\'\i sica Te\'orica, Universidade Estadual
Paulista}
\centerline{\it Rua Pamplona 145, 01405-900, S\~ao Paulo, SP, Brasil}

\vskip .3in
Using the U(4) hybrid formalism, manifestly N=(2,2) worldsheet
supersymmetric sigma models are constructed for the Type IIB superstring
in Ramond-Ramond backgrounds. The Kahler potential in these N=2 sigma models
depends on four chiral and antichiral bosonic superfields and two chiral
and antichiral fermionic
superfields. When the Kahler potential is quadratic, the 
model is a free conformal field theory which describes a flat
ten-dimensional target space with
Ramond-Ramond flux and non-constant dilaton.
For more general Kahler potentials, the model describes
curved target spaces with Ramond-Ramond flux that are not plane-wave
backgrounds. Ricci-flatness
of the Kahler metric implies the on-shell conditions for the background
up to the usual four-loop conformal anomaly.

\Date {October 2002}

\newsec{Introduction}

Because of the AdS/CFT correspondence \mone, understanding the superstring
in backgrounds with Ramond-Ramond (RR) flux is an important problem.
For plane-wave RR backgrounds, one can use the light-cone Green-Schwarz
(GS) formalism
to compute the physical spectrum \metsaev\bmn. 
In principle, the light-cone GS
formalism can also be used to compute scattering amplitudes, however,
there are complications caused by interaction-point operators
and contact terms. For RR backgrounds which do not allow a light-cone
gauge choice, one can use the classical covariant GS formalism
to study supergravity properties of the background \tseyt. However, it
is difficult to compute string-related properties of the background because
of quantization problems in the covariant GS formalism.

An alternative approach to study RR backgrounds is to use the
new covariant formalism involving pure spinors \pures. This formalism
manifestly preserves all isometries of the background and is
easy to quantize using a BRST operator. Unfortunately, although
it is straightforward to construct quantizable actions for
the superstring in $AdS_5\times S^5$ \meads\ and plane-wave RR backgrounds
\confops\ using this formalism, the resulting actions are non-linear and it is
not yet known how to simplify or solve them.

Recently \maluf, it has been realized that another approach called
the U(4) hybrid formalism \ufourp\ may be very useful for studying
Type IIB RR backgrounds. This U(4) hybrid formalism is manifestly
N=(2,2) worldsheet supersymmetric and contains four chiral
and antichiral bosonic superfields, $X^l$ and $X^\lb$ for $l=1$ to 4,
two chiral and antichiral fermionic superfields,
$(\T_L,\T_R)$ and $(\Tb_L,\Tb_R)$, and two semi-chiral
and semi-antichiral fermionic superfields,
$(W_L,W_R)$ and $(\Wb_L,\Wb_R)$. Although this formalism
only manifestly preserves 25 of the 45 SO(9,1) Lorentz
transformations, it is also manifestly invariant under 20 of the 32
Type IIB supersymmetries. As was shown in \topo, the formalism
is a critical N=2 superconformal field theory which is related
by a field redefinition to the N=1 $\to$ N=2 embedding of the
RNS superstring. And in light-cone gauge, where the fermionic superfields
are gauged away, the U(4) hybrid formalism reduces to the light-cone
GS formalism including the correct interaction-point operators \ufourp.

In \maluf, the U(4) hybrid formalism was used to describe plane-wave
RR backgrounds \maoz\ in which all non-zero RR field strengths have
a spacetime $+$ index. (Throughout this paper, the $\pm$ index
will always refer to the $x^\pm={1\over\sqrt{2}}
(x^9\pm x^0)$ spacetime directions.)
For plane-wave RR backgrounds, the U(4) worldsheet action 
depends trivially on the $(W,\Wb)$ superfields and closely resembles
the light-cone GS action \maoz. However, unlike the light-cone GS action, 
the U(4) worldsheet action does not require contact terms and 
quantum superconformal invariance of
the action implies the on-shell
conditions of the background.\foot{It was recently claimed that
any plane-wave RR background which satisfies the supergravity
equations at lowest order in $\a'$ is a consistent superstring
background\russotseytlin. In the U(4) formalism, this claim implies that
any plane-wave RR background which is superconformally invariant
at one-loop is superconformally invariant at all loops. This is
reasonable since, as was discussed in \maluf, plane-wave RR backgrounds involve
interaction vertices which are quadratic in $\T$. 
Since there are only four independent $\T$'s, any divergent 
counterterm can involve at most two interaction vertices. But above
one-loop, all known counterterms in N=2 sigma models involve more
than two vertices. For example, the well-known four-loop $R^4$
counterterm \grizanon\ involves a minimum of four vertices. So if one could
prove that all possible counterterms above one-loop in
N=2 sigma models involve more than two vertices, one will have
proven the claim of \russotseytlin. }

Although plane-wave RR backgrounds are interesting to study because
they are Penrose limits of the $AdS_5\times S^5$ background, it
would be useful to also have simple conformal field theory 
descriptions of more general RR backgrounds. As will be shown here,
the U(4) hybrid formalism can not only be used to describe 
backgrounds with RR field strengths containing a spacetime $+$ index,
but can also be used to describe
certain RR field strengths containing a spacetime $-$ index. Such
backgrounds do not allow a light-cone gauge choice and therefore
cannot be described using the light-cone GS formalism.

In the U(4) hybrid formalism, there are two special RR field strengths
containing a spacetime $-$ index whose vertex operators are
$\int d^2 z \int d^4\k ~W_L \Wb_R$
and $\int d^2 z \int d^4\k ~W_R \Wb_L$ where
$(W_R,W_L,\Wb_R,\Wb_L)$ are the semi-chiral and
semi-antichiral superfields. If these RR field strengths have
non-zero flux, the $W$ superfields satisfy auxiliary equations of
motion and can be integrated out of the worldsheet action. The resulting
worldsheet action is an N=(2,2) sigma model which depends on
four chiral and antichiral bosonic superfields, $X^l$ and $\Xb^\lb$,
and two chiral and antichiral fermionic superfields, $(\T_L,\T_R)$
and $(\Tb_L,\Tb_R)$, through a Kahler potential
$S=\int d^2 z \int d^4\k ~K(X,\Xb,\T,\Tb)$. The background is on-shell
up to the usual four-loop conformal anomaly \grizanon\ when the Kahler
metric $G_{M\Nb}=\p_M\overline\p_\Nb K$ is Ricci-flat \gaume. 

It is interesting to ask what types of 
superstring backgrounds are described by the Kahler potential
$K(X,\Xb,\T,\Tb)$. When the Kahler potential is quadratic,
i.e. $K=X^l\Xb^\lb +\T_L\Tb_R + \T_R\Tb_L$, it will
be argued that the background is a flat ten-dimensional spacetime
containing non-zero RR field strengths with a spacetime $-$ index
and a non-constant dilaton field $\phi$ which depends quadratically
on $x^-$. So this RR background with a non-constant dilaton is described
by a free conformal field theory. It would be interesting to construct
vertex operators and compute scattering amplitudes in this RR
background using the free-field OPE's implied by the action.

More general Kahler potentials will be shown to
describe certain curved backgrounds 
which contain non-zero RR field strengths with both
spacetime $+$ and $-$ indices. It should be possible to use
standard N=(2,2) superconformal methods to study the physical spectrum
and scattering amplitudes in these RR backgrounds which do not allow
a light-cone gauge choice. Hopefully, this information will be useful
for learning more about the AdS/CFT correspondence.

In section 2 of this paper, the U(4) hybrid formalism and its
field redefinition to the RNS formalism will
be reviewed in a flat background. In section 3, the RR vertex operators
$\int d^2 z d^4\k ~W_L\Wb_R$ and
$\int d^2 z d^4\k ~W_R\Wb_L$ will be added to the flat action and
the resulting linear N=2 sigma model will be discussed.
And in section 4, the properties of more general non-linear N=2 sigma models
using the U(4) hybrid formalism will be described.

\newsec{U(4) Hybrid Formalism in a Flat Background}

One way of understanding the U(4) hybrid formalism in a flat
background is as a field redefinition from the N=1$\to$N=2
embedding of the RNS formalism \topo. 
Recall that the critical N=1 description
of the RNS superstring
can be embedded into a critical N=2 description by defining the
$\hat c=2$ N=2 superconformal generators as
\eqn\generators{T= T_{N=1}^{matter}+ T_{N=1}^{ghost} +\half \p(bc+\xi\eta),}
$$G= j_{BRST},$$
$$\Gb = b,$$
$$J= bc +\xi\eta,$$
where $Q=\int dz ~j_{BRST}$ is the RNS BRST operator
and $(\xi,\eta)$ come from bosonizing the N=1 super-reparameterization
ghosts as $\beta=\p\xi e^{-\phi}$ and $\gamma=\eta e^\phi$.
Note that the term $\half (bc+\xi\eta)$ in $T$ shifts the conformal
weights of the RNS ghosts so that $G$ and $\Gb$ have
conformal weight $3\over 2$. As discussed in \topo, physical
states in the RNS formalism can be defined as N=2 superconformally
invariant states with respect to these N=2 generators, and RNS scattering
amplitudes can be computed using standard N=2 rules.

Although the RNS worldsheet variables transform in a complicated
non-linear manner under the N=2 superconformal transformations generated
by $[T,G,\Gb,J]$, one can define a field redefinition to U(4)
hybrid variables which transform linearly under these N=2
transformations. These left-moving U(4) hybrid variables are \ufourp\
\eqn\hybridv{[x^l,\xb^\lb, s_L^l,\sb_L^\lb,
p_L,\pb_L,\t_L,\tb_L, \l_L,\bl_L,w_L,\wb_L]
{\rm ~~for ~~}l=1
{\rm ~~to~~ }4 } 
and
are related to the left-moving RNS variables
$[x^\mu_{RNS},\psi^\mu,b,c,\xi,\eta,\phi]$ for $\mu=0$ to 9
by the field redefinition
\eqn\fieldredef{x^l = {1\over {\sqrt 2}}(x^l_{RNS}+i x^{l+4}_{RNS}),}
$$
\xb^\lb = {1\over {\sqrt 2}}(x^l_{RNS}-i x^{l+4}_{RNS} + c\xi e^{-\phi}
(\psi^l -i\psi^{l+4})),$$
$$s^l = {1\over{\sqrt 2}} \eta e^\phi (\psi^l +i\psi^{l+4})
+
{1\over{\sqrt 2}} c \p_L(x_{RNS}^l +i x_{RNS}^{l+4}), $$
$$
\sb^\lb = 
 {1\over{\sqrt 2}} \xi e^{-\phi} (\psi^l -i\psi^{l+4}),$$
$$p_L=
b\eta e^{{3\over 2}\phi}\Sigma^{-++++} + $$
$$ 
e^{\half \phi} (\p_L x^-_{RNS} \Sigma^{+++++}
+{1\over\sqrt{2}}
\p_L(x_{RNS}^l +i x_{RNS}^{l+4})\Sigma^{-\lb}) -\p_L(x^-_L e^{\half\phi}
\Sigma^{+++++}),$$
$$\pb_L= e^{-\half\phi}\Sigma^{-----} - \p_L(c x^-_L\xi e^{-{3\over 2}\phi}
\Sigma^{+----}),$$
$$\t_L = e^{\half\phi}\Sigma^{+++++},$$
$$\tb_L = c\xi e^{-{3\over 2}\phi}\Sigma^{+----},$$
$$\l_L=
 \eta e^{{3\over 2}\phi} (\p_L x^+_{RNS} \Sigma^{-++++}
+{1\over\sqrt{2}}
\p_L (x^l_{RNS} + i x^{l+4}_{RNS})\Sigma^{+\lb})  +$$
$$ c \p_L(e^{\half\phi}
\Sigma^{+++++}) + b\eta\p_L\eta e^{{5\over 2}\phi}\Sigma^{+++++},$$
$$\bl_L=\xi e^{-{3\over 2}\phi}\Sigma^{+-----},$$
$$w_L= Q( c\xi e^{-\half\phi}\Sigma^{-++++} + x_L^- e^{\half\phi}
\Sigma^{+++++}),$$
$$\wb_L = x_L^-\xi e^{-{3\over 2}\phi} \Sigma^{+----},$$
where 
$x_L^-$ and $x_R^-$ are the right and left-moving
contributions to 
\eqn\defxminus{x^-_{RNS}={1\over{\sqrt{2}}}(x^9_{RNS}-x^0_{RNS})= 
x^-_L + x^-_R,} 
$\Sigma^{\pm\pm\pm\pm\pm}$ are the 32 RNS spin fields constructed
by bosonizing $\psi^\mu$, and 
$$\Sigma^{+\lb}= (\Sigma^{+-+++},\Sigma^{++-++},\S^{+++-+},\S^{++++-}),$$
$$\Sigma^{-\lb}= (\Sigma^{--+++},\Sigma^{-+-++},\S^{-++-+},\S^{-+++-}).$$

Although the field redefinition of \fieldredef\ looks complicated, it
can be derived by first defining $x^l$ and $\xb^\lb$ such that
$b_0(x^l)=Q(\xb^\lb)=0$, which implies that $x^l$ and
$\xb^\lb$ are N=2 chiral and antichiral with respect to the generators
of \generators. One then defines
$s^l = Q(x^l)$ and $\sb^\lb= b_0(\xb^\lb)$. The next step is to define
$p_L$ and $\pb_L$ such that
$\int dz_L ~p_L$ and $\int dz_L~ \pb_L$ are the $q^{-++++}_{\half}$
and $q^{-----}_{-\half}$ supersymmetry
generators in the $+\half$ and $-\half$ picture of 
Friedan-Martinec-Shenker \fms, 
and then adds total derivative
terms to $p_L$ and $\pb_L$ so that they have non-singular OPE's
with each other. 
$\t_L$ and $\tb_L$ are then defined
to be conjugates to $\pb_L$ and $p_L$. Finally,
one defines $\l_L=Q(\t_L)$ and $\bl_L=b_0(\tb_L)$,
and defines $w_L$ and $\wb_L$ to be 
conjugates to $\bl_L$ and $\l_L$ which have
non-singular OPE's with the other fields. 
Similarly, the field redefinition for the right-moving U(4) hybrid
variables is obtained by replacing all left-moving fields
with right-moving fields
in \fieldredef.

In terms of the U(4) hybrid variables, one can check that the
left-moving N=2 generators of \generators\ are mapped to
\eqn\hybridgen{T_L= \p_L x^l\p_L \xb^\lb - \half(s_L^l\p_L\sb_L^\lb +
\sb_L^\lb \p_L s_L^l)}
$$-
\pb_L\p_L\t_L -p_L\p_L\tb_L +\half (w_L\p_L\bl_L-\bl_L\p_L w_L)
+\half(\wb_L\p_L\l_L - \l_L\p_L\wb_L),$$
$$G_L= s_L^l \p_L \xb^\lb + \l_L \pb_L + w_L\p_L\tb_L,$$
$$\Gb_L= \sb_L^\lb \p_L x^l + \bl_L p_L + \wb_L\p_L\t_L,$$
$$J_L= s_L^l \sb_L^\lb - w_L\bl_L +\wb_L\l_L,$$
and the RNS worldsheet action is mapped to 
\eqn\hybridaction{S= \int d^2 z [
\p_L x^l\p_R x^\lb -s_L^l\p_R\sb_L^\lb -s_R^l\p_L\sb_R^\lb}
$$ -
p_L\p_R\tb_L -\pb_L\p_R\t_L +w_L\p_R\bl_L
+\wb_L\p_R\l_L 
-p_R\p_L\tb_R -\pb_R\p_L\t_R +w_R\p_L\bl_R
+\wb_R\p_L\l_R].$$
So the U(4) variables of 
\hybridv\ satisfy free-field OPE's and
transform linearly under the N=2 superconformal
transformations generated by \hybridgen.

To construct these linearly transforming variables, it was necessary to
explicitly separate $x^-_{RNS}$ into its
left and right-moving parts as $x^-_{RNS}= x^-_L + x^-_R$. As
will now be discussed, this separation implies that physical
states in the U(4) hybrid formalism must not only be N=2
superconformally invariant, but must also satisfy an additional
global constraint. 

Using the field redefinition of \fieldredef, one can check that
\eqn\llconstraint{\l_L\bl_L -\tb_L\p_L\t_L = \p_L x^+_{RNS} {\rm ~~and~~}
\l_R\bl_R -\tb_R\p_R\t_R = \p_R x^+_{RNS} .}
Since $\oint dz_L ~\p_L x^+_{RNS}=
\oint dz_R ~\p_R x^+_{RNS}$ for any closed
contour, the U(4) hybrid variables satisfy the
global constraint
\eqn\glconstraint{\oint dz_L (\l_L\bl_L -\tb_L\p_L\t_L)-
\oint dz_R (\l_R\bl_R -\tb_R\p_R\t_R) =0.}
Using canonical commutation relations,
the constraint of \glconstraint\ generates 
the global isometry
\eqn\glisom{\d w_L=\a \l_L,\quad \d \wb_L=\a\bl_L, \quad
\d w_R=-\a \l_R,\quad \d \wb_R=-\a\bl_R, }
$$\d p_L=-\a \p_L\t_L,\quad \d \pb_L=-\a\p_L\tb_L, \quad
\d p_R=\a \p_R\t_R,\quad \d \pb_R=\a\p_R\tb_R.$$
And from \fieldredef, one can check that this isometry corresponds
in the RNS formalism to 
\eqn\xshift{\d x^-_L=\a,\quad \d x^-_R=-\a.}
So physical states in
the U(4) hybrid formalism must satisfy the global constraint of 
\glconstraint, which implies that they only depend on $x_L^-$
and $x_R^-$ in the combination $x_{RNS}^- = x_L^- + x_R^-$.

Since the U(4) hybrid variables transform linearly under N=(2,2)
worldsheet supersymmetry, it is convenient to combine them into
the N=(2,2) chiral and antichiral superfields
$$X^l (\k_L,\k_R)= x^l +\k_L s_L^l + \k_R s_R^l +
\k_L\k_R t^l,\quad
\Xb^\lb (\kb_L,\kb_R) = \xb^\lb +\kb_L \sb_L^\lb + \kb_R \sb_R^\lb + \kb_L
\kb_R \ttb^\lb ,$$
\eqn\xchiral{\T_L(\k_L,\k_R) =\t_L+ \k_L\l_L + ...,\quad
\T_R(\k_L,\k_R) =\t_R+ \k_R\l_R + ...,}
$$\Tb_L(\kb_L,\kb_R) =\tb_L+ \kb_L\bl_L + ...,\quad
\Tb_R(\kb_L,\kb_R) =\tb_R+ \kb_R\bl_R + ... ,$$
and the N=(2,2) semi-chiral and semi-antichiral superfields,
$$W_L(\k_L,\k_R,\kb_L) = ... +\k_L w_L -\k_L\kb_L p_L +  ...,\quad
W_R(\k_L,\k_R,\kb_R) = ...+ \k_R w_R -\k_R\kb_R p_R +  ...,$$
$$\Wb_L(\k_L,\kb_L,\kb_R) =...+ \kb_L \wb_L -\kb_L\k_L \pb_L +  ...,\quad
\Wb_R(\k_R,\kb_L,\kb_R) = ...+\kb_R \wb_R -\kb_R\k_R \pb_R +  ...,$$
where $t^l$, $\ttb^\lb$ and $...$ denote auxiliary fields 
which can be gauged away or vanish on-shell in a flat background.
Defining
\eqn\defderiv{D_L={\p\over{\p\k_L}}+\half \kb_L\p_L,\quad
D_R={\p\over{\p\k_R}}+\half \kb_R\p_R,}
$$\Db_L={\p\over{\p\kb_L}}+\half \k_L\p_L,\quad
\Db_R={\p\over{\p\kb_R}}+\half \k_R\p_R,$$
these superfields are constrained to satisfy the chirality constraints 
\eqn\chiralityc{\Db_L X^l = \Db_R X^l = D_L \Xb^\lb=D_R \Xb^\lb=0,}
$$\Db_L\T_L = \Db_R\T_L = \Db_L\T_R = \Db_R\T_R= 
D_L\Tb_L = D_R\Tb_L = D_L\Tb_R = D_R\Tb_R= 0,$$
$$\Db_R W_L = \Db_L W_R = D_R \Wb_L = D_L\Wb_R=0.$$

In terms of these N=(2,2) superfields, the N=2 left-moving
stress-tensor of \hybridgen\ is 
$${\cal T}_L=
(\Db_L \Wb_L) D_L\T_L - (D_L W_L) \Db_L\Tb_L + D_L X^l \Db_L\Xb^\lb,$$
the worldsheet action of \hybridaction\ is
\eqn\susyact{S = \int d^6 Z ~[
X^l\Xb^\lb + W_L\Tb_L + W_R\Tb_R + \Wb_L\T_L + \Wb_R\T_R],}
the global constraint of \glconstraint\ is
\eqn\susycon{\int d^3 Z_L  ~\T_L\Tb_L -
\int d^3 Z_R ~\T_R\Tb_R=0,}
and the isometry of \susycon\ is
\eqn\isotr{\d W_L = \a \T_L, \quad
\d W_R = -\a \T_R, \quad \d\Wb_L = \a\Tb_L,\quad\d\Wb_R = -\a\Tb_R,}
where $\int d^6 Z$ denotes
$\int d^2 z \Db_L\Db_R D_L D_R$, 
$\int d^3 Z_L$ denotes $\half\oint dz_L (\Db_L D_L - D_L \Db_L)$
and 
$\int d^3 Z_R$ denotes $\half\oint dz_R (\Db_R D_R -  D_R \Db_R)$.

With respect to SO(9,1) super-Poincar\'e transformations,
the U(4) hybrid
formalism is manifestly invariant under 20 of the 32 Type IIB spacetime
supersymmetries which are generated by
\eqn\susygen{q_L^{+l}=\int d^3 Z_L~ X^l\Tb_L,\quad
q_L^{+\lb}=\int d^3 Z_L ~\Xb^\lb\T_L,}
$$q_L^{-----}= \int d^3 Z_L ~W_L,\quad
q_L^{-++++}= \int d^3 Z_L ~ \Wb_L,$$
$$q_R^{+l}=\int d^3 Z_R ~X^l\Tb_R,\quad
q_R^{+\lb}=\int d^3 Z_R ~\Xb^\lb\T_R,$$
$$q_R^{-----}= \int d^3 Z_R ~W_R,\quad
q_R^{-++++}= \int d^3 Z_R ~\Wb_R,$$
under 25 of the 45 Lorentz transformations generated by
\eqn\lorentzgen{M^{l\mb} = \int d^3 Z_L ~X^l\Xb^\mb
+ \int d^3 Z_R  ~X^l\Xb^\mb, }
$$M^{+l} = \int d^3 Z_L ~X^l\T_L\Tb_L +
\int d^3 Z_R ~X^l\T_R\Tb_R ,$$
$$M^{+\lb} = \int d^3 Z_L ~\Xb^\lb\T_L\Tb_L +
 \int d^3 Z_R ~\Xb^\lb\T_R\Tb_R ,$$
$$M^{+-} = \int d^3 Z_L ~ (W_L\Tb_L -\Wb_L\T_L) +
\int d^3 Z_R ~(W_R\Tb_R -\Wb_R\T_R),$$
and under 9 of the 10 translations generated by
\eqn\transgen{P^l = -\int d^3 Z_L ~X^l
= -\int d^3 Z_R ~X^l , }
$$P^\lb = \int d^3 Z_L ~\Xb^\lb=
 \int d^3 Z_R ~ \Xb^\lb,$$
$$P^+ = \int d^3 Z_L~ \T_L\Tb_L=
\int d^3 Z_R~ \T_R\Tb_R.$$
Note that the translation generator $P^-$ does not act
on the U(4) hybrid variables since the field redefinition of
\fieldredef\ does not involve the zero mode of $x^+_{RNS}$.

In this paper, only the Type IIB version of the U(4) formalism
will be discussed. As was recently shown in \iiauf, treating the Type IIA
superstring in an N=(2,2) worldsheet supersymmetric manner
requires switching one of the four superfields 
$X^l$ from a chiral superfield to a twisted-chiral superfield which
breaks the manifest U(4) down to U(1) $\times$ U(3).
For the heterotic version of the U(4) formalism,
only N=(2,0) worldsheet supersymmetry is present and the right-moving
sector of the superstring is the same as in the RNS formalism.

\newsec{Linear N=2 Sigma Model}

In a recent paper with Maldacena \maluf, it was shown that the
U(4) hybrid formalism can be generalized to plane-wave backgrounds
in which the non-zero RR field strengths carry a spacetime $+$
index. These RR field strengths appear in the worldsheet action 
through the vertex operator
\eqn\rrplanev{\int d^6 Z [f_1(X,\Xb)\T_L\T_R + f_2 (X,\Xb)\Tb_L\T_R
-\fb_1(X,\Xb)\Tb_L\Tb_R -\fb_2 (X,\Xb)\T_L\Tb_R],}
which is constructed from left-right products of the
spacetime supersymmetry generators $(q^{+l}_L,q^{+\lb}_L)$ and
$(q^{+l}_R,q^{+\lb}_R)$ of \susygen.
In this section, the U(4) hybrid formalism will be generalized
to a background containing certain RR field strengths containing
a spacetime $-$ index. These field strengths will couple through
the vertex operator
\eqn\rrnewv{\int d^6 Z [ f W_L\Wb_R -\fb \Wb_L W_R],}
which is constructed from left-right products of the spacetime
supersymmetry generators $(q^{-----}_L,q^{-++++}_L)$ and
$(q^{-----}_R,q^{-++++}_R)$ of \susygen. Although $f$ and $\fb$
will be assumed to be constants in this paper, it might
be possible to consider more general RR vertex operators which depend
on both $W$ and $X$.

If one defines 
\eqn\bispinor{F^{\a\b}=F_{(1)}^\mu \g_\mu^{\a\b} +{1\over 6}
F_{(3)}^{\mu\nu\rho}
\g_{\mu\nu\rho}^{\a\b} + {1\over{480}}F_{(5)}^{\mu\nu\rho\sigma\tau}
\g_{\mu\nu\rho\sigma\tau}^{\a\b},}
where $[F_{(1)}^\mu,F_{(3)}^{\mu\nu\rho}, F_{(5)}^{\mu\nu\rho\sigma\tau}]$
are the Type IIB RR field strengths and $(\a,\b)=1$ to 16
are Majorana-Weyl spinor indices, then turning on the vertex operator
of \rrnewv\ corresponds to giving $e^\phi F^{\a\b}$ the background
value
\eqn\defff{e^\phi F^{\a\b} =\half M_\g^\a (\g^-)^{\g\d} \Mb^\b_\d f
+ 
\half \Mb_\g^\a (\g^-)^{\g\d} M^\b_\d \fb}
where $\phi$ is the dilaton and 
\eqn\defmm{
M_\a^\b = {1\over 4}
[(\g^1 + i\g^5)(\g^2 +i\g^6)(\g^3+i\g^7)(\g^4+i\g^8)]_\a{}^\b,}
$$\Mb_\a^\b = {1\over 4}
[(\g^1 -i\g^5)(\g^2 -i\g^6)(\g^3-i\g^7)(\g^4-i\g^8)]_\a{}^\b$$
are matrices which select the appropriate bispinor components of the
RR field strength.
So adding the vertex operator of \rrnewv\ corresponds to giving non-zero
flux to the components $F^{(1)}_-$, $F^{(3)}_{- l\lb}$, and
$F^{(5)}_{-lm\lb\mb}$ with relative coefficients
which depend on $f$ and $\fb$.

When $f$ and $\fb$ are constants, the vertex operator of \rrnewv\ is
$\int d^2 z  (f p_L\pb_R + \fb p_R\pb_L)$. So the worldsheet action
of \hybridaction\ or \susyact\
is still quadratic after adding this vertex operator.
In the presence of this vertex operator, the equations of motion for $W$ and
$\Wb$ become auxiliary and one can integrate them out \witten\ to obtain
the linear sigma model action
\eqn\intact{S=\int d^6 Z ~(X^l\Xb^\lb +\fb^{-1}\T_L\Tb_R + f^{-1}
\T_R\Tb_L)}
$$=\int d^2 z 
[\p_L x^l\p_R \xb^\lb +
\fb^{-1}(\p_R\t_L)(\p_L\tb_R)+f^{-1}(\p_L\t_R)(\p_R\tb_L) $$
$$-s_L^l\p_R\sb_L^\lb -s_R^l\p_L\sb_R^\lb +w_L\p_R\bl_L
+\wb_L\p_R\l_L +w_R\p_L\bl_R +\wb_R\p_L\l_R].$$
Note that $(\T_L,\Tb_L)$ 
and $(\T_R,\Tb_R)$ 
are
no longer left and right-moving functions on-shell. 
In components, 
$$\T_L(\k_L,\k_R) =\t_L+ \k_L\l_L +\fb\k_R  w_R + ...,\quad
\T_R(\k_L,\k_R) =\t_R+ \k_R\l_R + f\k_L  w_L + ...,$$
$$\Tb_L(\kb_L,\kb_R) =\tb_L+ \kb_L\bl_L - f\kb_R\wb_R + ... ,\quad
\Tb_R(\kb_L,\kb_R) =\tb_R+ \kb_R\bl_R -\fb \kb_L\wb_L + ... ,$$
where $...$ are auxiliary fields which vanish on-shell. 
The left and right-moving stress-tensors are now
\eqn\newstress{{\cal T}_L=
-\fb^{-1}(\Db_L\Tb_R )(D_L\T_L )-f^{-1}( D_L \T_R)( \Db_L\Tb_L )
+ D_L X^l \Db_L\Xb^\lb,}
$${\cal T}_R=
-f^{-1}(\Db_R\Tb_L )(D_R\T_R )- \fb^{-1}(D_R \T_L)( \Db_R\Tb_R)
 + D_R X^l \Db_R\Xb^\lb,$$
and are still quadratic in this RR background. 

Unlike the U(4) formalism in the plane-wave RR backgrounds discussed
in \maluf, the global constraint of \susycon\ needs to be modified
in this RR background since 
$(\T_L,\Tb_L)$ are no longer left-moving and
$(\T_R,\Tb_R)$ are no longer right-moving.
To find the correct
modification to \susycon, note that the action of \intact\ is invariant
under the global transformation
\eqn\glnew{\d\T_L = -\a \fb \T_R,\quad
\d\T_R = \a f \T_L,\quad
\d\Tb_L = \a f \Tb_R,\quad
\d\Tb_R = -\a \fb \Tb_L.} 
Furthermore, the auxiliary equations of motion for $W$ and $\Wb$
imply that they transform under 
\glnew\ in the same way as in \isotr.

Using the Noether method, the invariance under \glnew\ implies
that 
\eqn\conserved{J_L=\Db_L D_L(\T_L\Tb_L -\T_R\Tb_R),\quad
J_R=\Db_R D_R(\T_L\Tb_L -\T_R\Tb_R),}
is a conserved current satisfying $\p_R J_L + \p_L J_R=0$.
One can therefore restrict physical states to carry zero charge
with respect to this current, i.e. to satisfy the global constraint
\eqn\newsusycon{\int d^3 Z_L (\T_L\Tb_L -\T_R\Tb_R) +
\int d^3 Z_R (\T_L\Tb_L -\T_R\Tb_R)=0 .}
Or in components, 
\eqn\newcompocon{\oint dz_L 
(\l_L\bl_L+ f\fb w_L \wb_L -\tb_L\p_L\t_L +\tb_R\p_L\t_R)=}
$$
\oint dz_R (\l_R\bl_R +f\fb w_R \wb_R -\tb_R\p_R\t_R +\tb_L\p_R\t_L).$$
Note that this global constraint reduces to \glconstraint\
when $f=\fb=0$ and $(\t_{L/R},\tb_{L/R})$ are
left/right-moving. 

Since the worldsheet action of \intact\ is N=2 superconformal invariant
at the quantum level, an obvious question
is what on-shell supergravity background is it describing. From the
form of the action, it is clear that the metric is flat
and there is non-zero RR flux in the directions described in \defff. In
string gauge, the graviton equation of motion implies that 
\eqn\gravem{ e^{-2\phi}(R_{\mu\nu} - 2\nabla_\mu\nabla_\nu \phi)
+ (F^2)_{\mu\nu} =0.}
Using the RR field strength of \defff, the only non-zero component
of $(F^2)_{\mu\nu}$ is
$(F^2)_{--}= 4 e^{-2\phi} f\fb$. So the background satisfies
\gravem\ if $R_{\mu\nu}=0$,
$(F^2)_{--}= 4 e^{-2\phi} f\fb$, and
\eqn\dilatonem{\phi(x) =  f\fb (x^-)^2  + \phi_0.}
One can easily check
that this choice of $\phi$ and $F$ is a solution of all the supergravity
equations of motion in string gauge.
Note that $e^{-\phi}\to 0$ as $|x^-|\to\infty$, so the energy
density
of this classical supergravity solution is everywhere finite.
Furthermore, one can argue that
this solution is not affected by $\a'$ corrections to the
supergravity equations because of the inability to construct Lorentz-invariant
quantities out of $\phi$, $F$, and their derivatives. 

So even though the worldsheet action of \intact\ is free, it describes
a non-trivial supergravity background with RR flux and non-constant
dilaton. It would be interesting to construct vertex operators and
compute scattering amplitudes using this quadratic worldsheet action and
compare with analogous supergravity computations. 

Since the action of \intact\ is so simple, it seems reasonable to look
for a generalization of the field redefinition of \fieldredef\ in
this RR background with non-constant $\phi$. Although
it is not known how to construct RNS sigma model actions in RR backgrounds,
there exists an alternative hybrid formalism which
was developed with Gukov and Vallilo in \guk\ for describing compactifications
to two dimensions. When the eight-dimensional
compactification manifold is flat,
the worldsheet variables in this d=2 hybrid
formalism are almost the same as in the U(4) hybrid formalism and
consist of
\eqn\dtwohv{[x^+,x^-, x^l,\xb^\lb, s^l_L,s^l_R,\sb^\lb_L,\sb^\lb_R,
\t_L,\t_R,\tb_L,\tb_R, p_L,p_R,\pb_L,\pb_R, \sigma_L,\sigma_R,\rho_L,\rho_R]}
where $(\rho_L,\sigma_L)$ and $(\rho_R,\sigma_R)$ are left and right-moving
chiral bosons. So the only difference between the worldsheet variables
of \dtwohv\ and the U(4) hybrid variables is that 
$[x^+,x^-,\rho_L,\sigma_L,\rho_R,\sigma_R]$ is exchanged with
$[\l_L,\l_R,\bl_L,\bl_R,w_L,w_R,\wb_L,\wb_R]$. 

It was
shown in \guk\ how to
construct a sigma model action for RR backgrounds using
the d=2 hybrid formalism, so it should be possible
to find the field redefinition which maps the U(4) and d=2 hybrid
formalisms into each other in the background of \defff.
As in a flat background, this field redefinition will require splitting
$x^-$ into $x^-_L$ and $x^-_R$, and there will be a resulting 
global constraint on the U(4) variables. 
One should be able to verify that this global constraint is 
\newsusycon\ and that
the isometry of \glnew\ is generated by $\d x^-_L=\a$ and
$\d x^-_R= -\a$ as in \xshift. Also, it should
be possible to understand the non-constant
dilaton of \dilatonem\ as coming from a non-trivial Jacobian
in the field redefinition.

\newsec{Non-Linear N=2 Sigma Model}

The linear sigma model $S=\int d^6 Z~(X^l\Xb^\lb +\fb^{-1}\T_L\Tb_R
+ f^{-1}\T_R\Tb_L)$ of the previous section has an obvious
generalization to the non-linear sigma model
\eqn\nonlin{S=\int d^6 Z ~K(X,\Xb,\T,\Tb)}
where $K(X,\Xb,\T,\Tb)$ is the Kahler potential.\foot{One can also 
add to the non-linear sigma model of \nonlin\ the Fradkin-Tseytlin-like
term 
$\a' \int d^2 z
[\int d\k_L d\k_R ~\Phi(X,\T) R +
\int d\kb_L d\kb_R ~\Phib(\Xb,\Tb) \Rb ]$
where $\Phi(X,\T)$ and $\Phib(\Xb,\Tb)$ are chiral and antichiral
target-space superfields and $R$ and $\Rb$ are chiral and antichiral
worldsheet superfields which describe the N=(2,2) supercurvature.
In components, $R = ... + \k_R\k_L (r + i t)$ and
$\Rb = ... + \kb_R\kb_L (r - i t)$
where $r$ is the worldsheet curvature and $t$ is the U(1) field strength
for worldsheet R-symmetry \gris. So this term contains the usual
$\a'\int d^2 z ~\phi r$ coupling where $\phi= \Phi+\Phib$ is the
spacetime dilaton superfield.}
In order to describe a consistent superstring background, this non-linear
sigma model 
must be N=(2,2) superconformal invariant at the quantum level and
must be invariant under a global isometry analogous to \glnew.

Although it would be very interesting to study the most general action
which satisfies these conditions, we shall restrict our attention in
this paper to the action
\eqn\genact{S = \int d^6 Z [k(X,\Xb) +  a(X,\Xb)\T_L\T_R +
\ab(X,\Xb)\Tb_R\Tb_L}
$$ + b(X,\Xb) (\sffb\T_L\Tb_R +\sfbf\T_R\Tb_L)+
 d(X,\Xb)\T_L\Tb_L\T_R\Tb_R],$$
which is invariant under the isometry of \glnew,
\eqn\gltwo{\d\T_L =-\a \fb \T_R,\quad
\d\T_R = \a f \T_L,\quad
\d\Tb_L = \a f \Tb_R,\quad
\d\Tb_R = -\a \fb \Tb_L.} 
So physical states in this background must be N=2 superconformal
invariant and must satisfy the global constraint
\eqn\glthree{\int d^3 Z_L \sqrt{f\fb} b(X,\Xb) (\T_L\Tb_L -\T_R\Tb_R) +
\int d^3 Z_R \sqrt{f\fb} b(X,\Xb) (\T_L\Tb_L -\T_R\Tb_R)=0 ,}
which generates the isometry of \gltwo.
As will now be shown,
the action of \genact\ describes a background which includes both
RR fluxes with a spacetime $+$ index that appear in plane-wave
backgrounds and RR fluxes with a spacetime $-$ index that
were discussed in the previous section. 

To learn what superstring background is described by \genact,
it is useful to put back the $(W,\Wb)$ dependence in the action.
Defining the isometry transformation of $W$ and $\Wb$ 
as in the previous sections, i.e. 
\eqn\isotwo{\d W_L = \a \T_L, \quad
\d W_R = -\a \T_R, \quad \d\Wb_L = \a\Tb_L,\quad\d\Wb_R =- \a\Tb_R,}
the term
\eqn\addterm{\int d^6 Z [W_L\Tb_L + W_R\Tb_R + \Wb_L\T_L + \Wb_R\T_R
+ f W_L\Wb_R + \fb W_R\Wb_L]}
is invariant under the isometry of \gltwo\ and \isotwo. 
After adding \addterm\ to the action
of \genact\ and integrating out $W$ and $\Wb$, the only effect
is to shift
$b(X,\Xb) \to b(X,\Xb) + {1\over{\sqrt{f\fb}}}$. So
\genact\ is equivalent to
\eqn\genacttwo{S = S_0 + \int d^6 Z [\kh (X,\Xb) +  a(X,\Xb)\T_L\T_R +
\ab(X,\Xb)\Tb_R\Tb_L +
f W_L\Wb_R + \fb W_R\Wb_L}
$$ + \bh(X,\Xb) (\sffb\T_L\Tb_R +\sfbf\T_R\Tb_L)+
d(X,\Xb)\T_L\Tb_L\T_R\Tb_R]$$
where $S_0$ is the action of \susyact\ in a flat background, 
\eqn\defhatted{\kh(X,\Xb)=
k(X,\Xb)-X^l\Xb^\lb\quad {\rm ~~and~~}\quad \bh(X,\Xb)=b(X,\Xb)- 
{1\over\sqrt{f\fb}}.}

By comparing with the massless vertex operators in a flat background,
one can easily determine the linearized values of the 
supergravity background fields
which contribute to \genacttwo. One finds  
\eqn\linorder{g_{l\mb}=\d_{l\mb}+\p_l\bp_\mb \kh, \quad g_{++}=-d +
(\bp_\lb a)(\p_l \ab),\quad g_{+-}=1,}
$$ \phi(x)=\phi_0 +
{1\over 4}(x^-)^2 f\fb,$$
$$e^\phi
F^{\a\b} = \half(\g^l)^{\a\g} \Mb_\g^\d \g^+_{\d\k} 
(\g^m)^{\k\b}
\nabla_l\nabla_m \ab 
+ 
\half(\g^\lb)^{\a\g} M_\g^\d \g^+_{\d\k}  
(\g^\mb)^{\k\b}
\nablab_\lb\nablab_\mb a $$
$$+{1\over 8}(\g^\lb)^{\a\g} M_\g^\d \g^+_{\d\k} \Mb^\k_\rho 
(\g^m)^{\rho\b}
\sffb \nablab_\lb \nabla_m \bh 
+{1\over 8} (\g^l)^{\a\g} \Mb_\g^\d \g^+_{\d\k} M^\k_\rho 
(\g^\mb)^{\rho\b}
\sfbf \nabla_l \nablab_\mb \bh$$
$$+ \half
 M_\d^\a (\g^-)^{\d\k} \Mb^\b_\k f
+ \half
\Mb_\d^\a (\g^-)^{\d\k} M^\b_\k \fb $$
where $(g_{l\mb}, g_{++},g_{+-})$ are the non-zero components of the metric,
$F^{\a\b}$ are components of the RR field strength written
in the bispinor notation of \bispinor,
$(\g^l)^{\a\b}= e^l_c (\g^c)^{\a\b}$ and
$(\g^\lb)^{\a\b}= e^\lb_c (\g^c)^{\a\b}$, $c=0$ to 9 is
a tangent-space vector index, 
$(e^l_c,e^\lb_c, e^+_c, e^-_c)$ is the vielbein satisfying
$e^l_c e^\mb_d \eta^{cd} =g^{l\mb}$ and $e^l_c e^m_d \eta^{cd} = 
e^\lb_c e^\mb_d \eta^{cd}=0$,
and $M_\a^\b$ and $\Mb_\a^\b$ are defined in \defmm.

Since the background values of \linorder\ have been determined from
infinitesimal vertex operators, they are only guaranteed to be
correct to linearized order in the fields 
\eqn\fqq{[\kh, a, \ab, \bh,d,f,\fb].}
But certain backreactions which are
quadratic in these fields can be determined from other analysis.
For example, the quadratic term $(\bp_\lb a)(\p_l\ab)$
in $g_{++}$ comes from integrating out the auxiliary variables
$t^l$ and $\ttb^\lb$ of \xchiral\ in a plane-wave
background.
And the quadratic
dependence on $f$ in $\phi(x)$ comes from the solution
of \dilatonem\ in a pure RR background. However, there are certainly
other backreactions which are cubic or higher-order in the fields of \fqq\
and which have been neglected in \linorder. So in the following 
analysis,
the background values of \linorder\ will be assumed to be correct only
up to quadratic order in the fields of \fqq. 
For example, the $f\fb (x^-)^2$ dependence in $\phi$ will be neglected
when considering the background value of $e^\phi F^{\a\b}$ in
the following analysis since this dependence only affects terms which
are at least cubic order in the fields of \fqq.

It will now be shown that up to quadratic order
in \fqq, N=2 superconformal
invariance implies that the background fields in \linorder\
satisfy the Type IIB supergravity equations of motion.
Note that 
superconformal invariance of an N=2 sigma model implies up
to a four-loop anomaly \grizanon\ that the Kahler metric is 
Ricci-flat \gaume,
i.e. 
\eqn\flat{\p_M\bp_\Nb [\log sdet (\p\bp K)]=0}
where $\p$ denotes derivatives 
with respect to
$(X^l,\T_L,\T_R)$ chiral superfields
and $\bp$ denotes derivatives with respect to 
$(\Xb^\lb,\Tb_L,\Tb_R)$ antichiral superfields.
Defining $K$ to be the Kahler potential of \genact\ where $\kh=
k-X^l\Xb^\lb$ and $\bh=b-{1\over{\sqrt {f\fb}}}$, 
one finds that 
\eqn\superdet{
sdet(\p\bp K)=f\fb ~det g ~[1 -2
\sqrt{f\fb}\bh }
$$+ \T_L\T_R g^{l\mb}\p_l\bp_\mb a 
+ \Tb_R\Tb_L g^{l\mb}\p_l\bp_\mb \ab 
+(\sqrt{ f\over\fb}\T_L\Tb_R+\sqrt{\fb\over f}\T_R\Tb_L)
(g^{l\mb}\p_l\bp_\mb \bh + \sqrt{f\fb} d)$$
$$+\T_L\Tb_L\T_R\Tb_R (g^{l\mb}\p_l\bp_\mb d 
-(\p_l\bp_\mb a)
(\bp_\lb\p_m \ab ) +(\p_l\bp_\lb a)(\p_m\bp_\mb\ab) $$
$$
+(\p_l\bp_\mb \bh )(\bp_\lb\p_m \bh )-
(\p_l\bp_\lb \bh)(\p_m\bp_\mb\bh))
 + ...]$$
where $g^{l\mb}\p_l\p_\mb=
\p_l\p_\lb -(\p_m\bp_\lb\kh)\p_l\bp_\mb$
and $...$ is at least cubic order in the fields of \fqq. So
up to terms quadratic order in these fields,  
\flat\ implies the equations
\eqn\impleqone{
\p_m\bp_\nb(\log det ~g
-2\sqrt{f\fb}\bh )=0,}
$$\bp_\nb(g^{l\mb} \p_l\bp_\mb a)=0,$$
$$\p_n(g^{l\mb}\p_l\bp_\mb \ab )=0,$$
$$g^{l\mb}\p_l\bp_\mb \bh +  \sqrt{f\fb} d=0,$$
$$g^{l\mb}\p_l\bp_\mb d -
(\p_l\bp_\mb a)
(\bp_\lb\p_m \ab ) 
+(\p_l\bp_\mb \bh )(\bp_\lb\p_m \bh ) =0.$$

It will now be shown that \impleqone\ implies that the background fields of
\linorder\ 
satisfy the Type IIB supergravity equations in string gauge. When
the NS-NS $B_{\mu\nu}$ field vanishes, the Type IIB supergravity
equations can be written in terms of the bispinor $F^{\a\b}$ of \bispinor\ as
\eqn\seqa{e^{-2\phi} (R_{\mu\nu}-2\nabla_\mu\nabla_\nu\phi)=
-{1\over {64}}\g_\mu^{\a\b}\g_\nu^{\g\d} F_{\a\g} F_{\b\d}
+ {g_{\mu\nu}\over {64}}F^{\a\b} F_{\a\b},}
\eqn\seqb{(\g_{\mu\nu})^\a{}_\g F^{\g\b} F_{\a\b}=0,}
\eqn\seqc{R=-4\nabla_\mu\phi \nabla^\mu\phi + 4\nabla^\mu\nabla_\mu\phi,}
\eqn\seqd{\g^\mu_{\a\b} \nabla_\mu F^{\b\g}=
\g^\mu_{\a\b} \nabla_\mu F^{\g\b}= 0.}
Note that \seqb\
comes from varying $B^{\mu\nu}$ in the supergravity action and
then setting $B^{\mu\nu}=0.$

To verify that these supergravity equations are implied by
\impleqone, first note
that the matrices $M_\a^\b$ and $\Mb_\a^\b$ of \defmm\ satisfy
\eqn\msatisfy{(\g^m)^{\a\b} M_\b^\g = 
(\g^\mb)^{\a\b} \Mb_\b^\g = M_\a^\b (\g^\mb)_{\b\g} M^\g_\d =
\Mb_\a^\b (\g^m)_{\b\g} \Mb^\g_\d =
0 ,}
which implies that the background value of
\linorder\ for $F^{\a\b}$ 
satisfies 
\eqn\fsatisfy{F^{\a\b} F_{\a\g} = F^{\a\b} F_{\g\b} =0,}
and that the only non-zero components of  
$e^{2\phi} \g_\mu^{\a\g}\g_\nu^{\b\d} F_{\a\b} F_{\g\d}$ are
\eqn\ffterms{e^{2\phi} \g_+^{\a\g}\g_+^{\b\d} F_{\a\b} F_{\g\d}=
64 (\p_l\p_m \ab)(\bp_\lb\bp_\mb a) +
64 (\p_l\bp_\mb \bh ) (\p_m\bp_\lb \bh ), }
$$e^{2\phi} \g_-^{\a\g}\g_-^{\b\d} F_{\a\b} F_{\g\d}= 256 f\fb,$$
$$e^{2\phi} \g_l^{\a\g}\g_\mb^{\b\d} F_{\a\b} F_{\g\d}= 
128 \sqrt{f\fb}\p_l\bp_\mb \bh.$$
Also, the background value for $g_{\mu\nu}$ of \linorder\ implies that the 
Ricci tensor $R_{\mu\nu}$ satisfies
\eqn\rrterms{R_{++}= - g^{l\mb}\p_l\bp_\mb g_{++} = 
g^{l\mb}\p_l\bp_\mb( d - (\bp_\nb a)(\p_n\ab)),}
$$R_{l\mb}= -\p_l\bp_\mb (\log det ~g),\quad R_{--}= R_{lm}=
R_{\lb\mb}=0.$$

Putting together \rrterms\ and \impleqone, one finds that
\eqn\findone{R_{++}= -(\p_l\bp_\mb \bh)(\bp_\lb \p_m \bh) -
(\p_l\p_m \ab)(\bp_\lb\bp_\mb a) {\rm ~~~and ~~~}
R_{l\mb}= -2\sqrt{f\fb} \p_l\bp_\mb \bh.}
So \seqa\ is satisfied using the values of \ffterms\ and \dilatonem\ for
$(F F)_{\mu\nu}$ and $\phi$.
Furthermore, \seqb\ is implied by \fsatisfy, and \seqc\ is implied
(up to quadratic order in \fqq) by \findone\ and the fourth
equation of \impleqone. 

Finally, \seqd\ can be verified using the background value of \linorder\
for $F^{\a\b}$ together with the identities $\g^{lm} \nabla_l\nabla_m = 
\g^{\lb\mb} \nablab_\lb\nablab_\mb = 0$, $\{\g^l,\g^\mb\}=
2 g^{l\mb}$, and \msatisfy\ to obtain 
\eqn\deriva{\g^\mu_{\a\b}\nabla_\mu F^{\b\g}=
(\g^l_{\a\b}\nabla_l +
\g^\lb_{\a\b}\nablab_\lb + \g^-_{\a\b}\nabla_-  + \g^+_{\a\b}\nabla_+)
F^{\b\g}}
$$= M_\a^\d \g^+_{\d\k} (\g^\mb)^{\k\g}\nabla_l\nablab_\lb\nablab_\mb a
+\Mb_\a^\d \g^+_{\d\k} (\g^m)^{\k\g}\nablab_\lb\nabla_l\nabla_m\ab$$
$$+{1\over 4}  
M_\a^\d \g^+_{\d\k} \Mb_{\k\rho}(\g^m)^{\rho\g}\sffb
\nabla_l\nablab_\lb\nabla_m\bh
+{1\over 4}  
\Mb_\a^\d \g^+_{\d\k} M_{\k\rho}(\g^\mb)^{\rho\g}\sfbf
\nablab_\lb\nabla_l\nablab_\mb\bh$$
$$+{1\over 4} \g^+_{\a\b} w_+{}^{cd}[(\g_{cd})^\b{}_\d F^{\d\g}
+(\g_{cd})^\g{}_\d F^{\b\d}]$$
where
\eqn\spinc{{1\over 4} w_+{}^{cd}(\g_{cd})^\b{}_\d =
{1\over 4}[\p_l g_{++} (\g^+\g^l)^\b{}_\d +\bp_\lb g_{++} 
(\g^+\g^\lb)^\b{}_\d]}
$$=-{1\over 4}
[\p_l d (\g^+\g^l)^\b{}_\d +\bp_\lb d
(\g^+\g^\lb)^\b{}_\d]$$
is the $+$ component of the spin connection to linearized order in
\fqq. Using \impleqone\ and 
\eqn\usefi{\g^+_{\a\b} (\g^+\g^l)^\g{}_\d F^{\b\d} = -
M_\a^\sigma \g^+_{\sigma\k} \Mb^\k_\rho (\g^l)^{\rho\g}f,}
$$\g^+_{\a\b} (\g^+\g^\lb)^\g{}_\d F^{\b\d} = -
\Mb_\a^\sigma \g^+_{\sigma\k} M^\k_\rho (\g^\lb)^{\rho\g}\fb,$$
one learns that 
$\g^\mu_{\a\b}\nabla_\mu F^{\b\g}=0$. Similarly, one can show that
$\g^\mu_{\a\b}\nabla_\mu F^{\g\b}=0$. 

So it has been verified up
to quadratic order in the fields of \fqq\ that N=2 superconformal
invariance of \genact\ implies the Type IIB supergravity equations
of motion for the background. It would be useful to determine the
complete backreaction of the background values in \linorder\ and verify this
to all orders in the fields of \fqq.
It would also be interesting to
consider more general actions than \genact\ and determine what is
the most general Type IIB supergravity background that can be
described as an N=2 sigma model using the U(4) hybrid formalism.

\vskip 15pt
{\bf Acknowledgements:}
I would like to thank Marcus Grisaru, Hirosi Ooguri, Cumrun Vafa, and
especially Juan Maldacena for useful discussions and
CNPq grant 300256/94-9, Pronex grant 66.2002/1998-9, 
and FAPESP grant 99/12763-0 for partial financial support.
This research was partially conducted during the period that NB
was employed by the Clay Mathematics Institute as a CMI Prize Fellow.

\listrefs

\end